# A new explanation to the cold nuclear matter effects in heavy ion collisions

Zhi-Feng Liu

(Institute of Geophysics, Shijiazhuang Economic University, Shijiazhuang, China)

The J/Psi cross section ratios of p-A/p-p under different collision energy is calculated with cold nuclear matter effects redifined in this paper. The advantage of these new definitions is that all cold nuclear matter effects have clear physical origins.The radios are compared with the corresponding experiment data and that calculated with classic nuclear effects. The ratios calculated with new definitions can reproduce almost all existing J/Psi measurements in p-A collisions more accurately than that calculated with classic nuclear effects. Hence, this paper presents a new approach to explain cold nuclear effects in the hardproduction of quarkonium.

*Supported by National Nature Science Foundation of China (10575028).

E-mail: liuzhifeng_cc@163.com.

# 1 Introduction

The quarkonium production in heavy ions collisions, such as the production of J/Psi, is a powerful probe of the properties of dense QCD matter (A-A) and a vital testing ground for novel nontrivial QCD dynamics (p-A). By studying the medium-induced modification of quarkonium production in nucleus collisions (p-A) relative to the naive proton-proton collisions (p-p), one can get information on the novel nontrivial QCD dynamics, especially with the process in d-Au collisions at the RHIC and the forthcoming p-Pb collisions at the LHC. The manifestation of such nontrivial QCD dynamics in p+A reactions is usually named as cold nuclear matter (CNM) effects-those effects cause an obviously suppression of cross section but not because of the formation of quark-gluon plasma. .

There have been several approaches to study CNM effects in the past decade. Based on the perturbative QCD factorization and fixed the parametrizations through a global fitting procedure, Eskola and Hirai et al attributes all these effects to universal nuclear parton distribution functions (nPDFs), where those nPDFs become the only ingredients different from the case of p-p collisions [1–2]. This approach can describe part of the RHIC d-Au data reasonably well. Beside them, Venugopalan et al present another approach named the so-called Color Glass Condensate (CGC) approach [3]. Because it focuses on non-linear corrections to QCD evolution equations in very dense gluonic systems and is only applicable in the very small-x region, it is not as popular as the first one. Another person, HIJING, had calculated the nuclear modification factor with Monte Carlo models.

In this paper we followed a new approach that derived from the multiple parton scattering theory. This approach is based on perturbative QCD factorization. At the same time CNM effects are implemented separately within the formalism by modification to the corresponding variables. The advantage of this approach is that all CNM effects have clear physical origin and mostly centered on the idea of multiple parton scattering [4]. This approach first derived in [5]. In Ref. [5] the cross section data of prompt photon in heavy ion collisions at RHIC was reproduced perfectly with this approach (see Fig1, Fig2). In their calculation the isospin effect, Cronin effect, cold nuclear matter energy loss, and dynamical shadowing were considered.

In this manuscript we will improve their model and calculate the production cross section ratio of J/Psi in heavy ion collisions. As we known, according to the multiple parton scattering theory, under some approximate conditions, at sufficiently large quarkonium energy E in the target rest frame, quarkonium hadroproduction looks like small angle scattering of a color charge [6]. Based on this theory, we used a new cold nuclear matter energy loss model derived from the medium–induced gluon radiation spectrun (similar to Bethe-Heitler spectrum) to calculate the fractional energy loss $\varepsilon$ [6], which is different from the effective reduced fractional energy loss $\varepsilon_{eff}$ used in [5]. Because our calculation of the $\varepsilon$ has included the effects of the transverse momentum nuclear broadening, we don't discuss the Cronin effect separately in the follow section. In fact, considering the high collision energy in heavy ion collisions, which induced that partons undergoing hard collisions are in a very small-x region (for example, about $10^{-4}$ at LHC), we add a gluon saturation effect to the calculation in the so-called classical gluon saturation regime [7]. In the following section 2 new CNM effects is introduced respectively. In section 3 the conclusion and figures are presented.

## 2 Cold nuclear matter effects

In general, the CNM effects in p-A collisions can be quantified with a nuclear modification factor $R_{pA}$, which is usually defined as:

$$R_{pA} = \frac{1}{A}\frac{d\sigma_{pA}}{dyd^2p_T} / \frac{d\sigma_{pp}}{dyd^2p_T} . \qquad (1)$$

Here A is the nucleon number of target nuclear. The deviation of $R_{pA}$ from unity reveals the presence of CNM effects in p-A collisions. In this paper we mainly discuss such CNM effects as the cold nuclear matter energy loss, gluon saturation effect and dynamical shadowing effect, for those effects have been theoretically evaluated as arising from the elastic, inelastic and coherent scattering of partons in large nuclei [4].

### 2.1 Cold nuclear matter energy loss

Cold nuclear matter energy loss effect can be easily implemented as a momentum fraction shift in the PDFs of the incoming proton. According to the multiple parton scattering theory, as the parton from the proton undergoes multiple scattering in the nucleus before hard collisions, it can lose energy due to medium-induced gluon bremsstrahlung, which correct the momentum fraction in the PDFs of the incoming proton with a shift $\varepsilon$ [5]:

$$f_{q/p}(x_a, Q^2) \to f_{q/p}(\frac{x_a}{1-\varepsilon}, Q^2), f_{g/p}(x_a, Q^2) \to f_{g/p}(\frac{x_a}{1-\varepsilon}, Q^2). \qquad (3)$$

Here $\varepsilon = \left\langle \sum_i \omega_i / E \right\rangle$ is the average fractional energy loss induced by multiple gluons emission. The sum runs over all medium-induced gluons. Ideally, Eq. (3) should include a convolution over the probability distribution of cold nuclear matter energy loss $P_{q,g}(\varepsilon)$ [8]. The probability distribution $P_{q,g}(\varepsilon)$ is calculated for quarks and gluons to lose a fraction of their energy $\varepsilon$ due to multiple gluon emission in the Poisson approximation [9]. This calculation of initial-state cold nuclear matter energy loss has been shown to give a good description of the nuclear modification of Drell-Yan production in fixed target experiments [10].

Instead of using an effective reduced fractional energy loss $\varepsilon_{eff}$ as in Ref. [5], the calculation of $\varepsilon$ in this paper follows the method in Ref. [6]. In their paper they assumed that the heavy-quark $Q\bar{Q}$ pair is produced in a compact color octet state, within the hard process time-scale $t_h$, and remains color octet for a time much longer than $t_h$. In quarkonium production models where color neutralization is a soft non-perturbative process, this assumption holds at any $x_F$. With this assumption, at sufficiently large quarkonium energy E in the target rest frame, quarkonium hadroproduction looks like small angle scattering of a color charge [6]. So the medium-induced initial-state radiation spectrum is similar to the Bethe-Heitler spectrum and is written as follow [6]:

$$\omega \frac{dI}{d\omega} = \frac{N_c \alpha_s}{\pi} \{\ln(1 + \frac{\Delta q_\perp^2 E^2}{M^2 \omega^2}) - \ln(1 + \frac{\Lambda^2 E^2}{M^2 \omega^2})\} \cdot \quad (4)$$

Here $\Delta \hat{q}_\perp^2$ is the transverse momentum nuclear broadening which is a variable related to the Cronin effect. M = 3GeV is the mass of a compact $c\bar{c}$ pair. The average initial-state radiative energy loss can be obtained by integrating the initial-state medium-induced bremsstrahlung spectrum [11]:

$$\Delta E \equiv \int_0^E d\omega \omega \frac{dI}{d\omega} \approx N_c \alpha_s \frac{\sqrt{\hat{q}_A(x)L}}{M_\perp} E \cdot \quad (5)$$

The average path length L is given by $L = \frac{3}{2} r_0 A^{1/3}$. $\hat{q}_A(x)$ stands for the transport coefficient in the nucleus A and is related to the gluon distribution $G(x)$ in the target nucleon as [12]:

$$\hat{q}(x) = \frac{4\pi^2 \alpha_s(\hat{q}L) N_c}{N_c^2 - 1} \rho x G(x), \quad \hat{q}L \approx \frac{4\pi^2 \alpha_s N_c}{N_c^2 - 1} \rho x G(x) \cdot \quad (6)$$

Here $\rho$ is the target nuclear density. $\alpha_s$ is the running parameter. $N_c$ is the number of color.

In the present study we have assumed for simplicity that the octet $Q\bar{Q}$ pair arises dominantly from the splitting of an incoming gluon. This should be a valid assumption for all p–A data considered in this paper, except at very large values of $x_F$ ($x_F > 0.8$), where quark-induced processes come into play.

The energy loss model used in this paper is based on multiple parton scattering theory and is calculated with an integration of the initial-state medium-induced bremsstrahlung spectrum. It has a clear physical origin that is different from the classic energy loss model [13].

**2.2 Dynamical shadowing effect**

Power-suppressed resumed coherent final-sate scattering of the struck partons leads to suppression of the cross section in the small-x region, which is named as dynamical shadowing effect. This effect represents the interactions between the scattering partons and the nuclear background chromo-magnetic field. So analogous to the generation of dynamical mass for electrons propagating in a strong electro-magnetic field, the dynamical shadowing effect can be interpreted as a generation of dynamical parton mass in the background gluon field of the nucleus and contribute to the cross section at the power corrections level with a modification to Bjorken $x_b$ of the target nucleus[8]. Such correction lead to a suppression of the single and double inclusive hadron production cross sections at forward rapidity as long as the coherence criterion is satisfied.

The dynamical parton mass, which is calculated in QCD, is given by $m_{dyn}^2 = \zeta^2(A^{1/3}-1)$ [14]. Then the corresponding change in the value of Bjorken $x_b$ is

$$x_b \to x_b(1+\frac{m_{dyn}^2}{Q^2}). \tag{7}$$

In the case of the production of J/Psi, the nuclear size enhanced ($A^{1/3}$) power correction can be resummed for a given partonic channel (t-channel) and lead to the following shift in the momentum fraction $x_b$ for the parton inside the nucleus[6]:

$$x_b \to x_b(1+C_d \frac{\xi^2(A^{1/3}-1)}{-\hat{t}}). \tag{8}$$

Here $x_b$ is the parton momentum fraction inside the target nucleus, $C_d = C_F(C_A)$ if the parton $d = q(g)$ in the partonic scattering $ab \to cd$. $\xi^2$ represents a characteristic scale of the multiple scattering per nucleon[15]. For $x_b < 0.1$, due to coherence, $m_{dyn}^2$ has to be incorporated in the underlying kinematics of the hard scattering in Eq. (7); for $x_b > 0.1$, because the vector meson exchange is localized to one nucleon, the elastic final-state interactions vanish.

Similar shifts in the momentum fraction $x_b$ can be obtained easily from Eq. (8) by substituting $\hat{t} \to \hat{u}$ or $\hat{t} \to \hat{s}$ respectively in other partonic channels [16].

All in a word, the dynamical shadowing effect named in this paper represents the interactions between the scattering outgoing partons and the nuclear background chromo-magnetic field, which leads to a shift in the momentum fraction $x_b$ inside the target nucleus. Comparing with the classic shadowing effect defined as the ratio of PDFs in nucleus vs. PDFs in proton [1, 2], the dynamical shadowing effect has more clear physical origin.

**2.3  Gluon saturation effect**

Because the parton momentum fraction x probed in collisions at RHIC and LHC is very small, we must consider the gluon saturation effect in our calculation. At small value of x, parton wave function inside the nucleus starts to overlap, which changes the effective collision number in nucleus and leads to an additional J/Psi suppression in high-energy p–A collisions. This phenomenon is named as gluon saturation effect. The suppression here is independent of that made by energy loss effect discussed above. In fact, because the saturation effect is expected to scale roughly as the nucleus transverse density, $V/S \sim A^{1/3}$, the J/Psi normalized yield in p–A collisions is likely to be suppressed with respect to that in p–p collisions either at large $x_F$ and/or at high energies.

In the present paper, we shall implement the physics of saturation following the work of Fujii, Gelis and Venugopalan [7], where J/ψ suppression has been calculated within the Color Glass Condensate assuming $2 \to 1$ kinematics for the J/Psi production process. The nuclear suppression is a scaling function of the saturation scale $Q_s$ and is simply parameterized as follow:

$$S_A^{J/\psi}(x_2, L) \approx (1 + \frac{Q_s^2(x_2, L)}{b})^{-\alpha}. \tag{9}$$

Here b = 2.65 GeV$^2$ and α = 0.417. The saturation scale $Q_s$ is determined through the relationship [17]:

$$Q_s^2(x, L) = \hat{q}(x) L. \tag{10}$$

$\hat{q}(x)$ is the transport coefficient. L is the average path length. Both of them have been presented in above section 2.1.

**3  Conclusion**

In order to verify the availability of the new cold nuclear matter effect model on explaining the production suppression in heavy ion collisions, the cross section ratios of J/Psi under different collision energy were calculated in this paper. The new model is based on the multiple parton scattering theory and includes such cold nuclear matter effects as the energy loss effect, isospin effect, Cronin effect, dynamical shadowing effect and gluon saturation. The results calculated with new model were compared with the experiment data collected under different collision energy as well as the results calculated with classic nuclear effect model. In the classic nuclear effect model, we calculated energy loss effect with GM model [12] and shadowing effects with EPS09 [1] and final-state nuclear absorption effect with Glauber model [18].

In Fig.1 the rapidity distribution of nuclear modification factor $R_{Au/p}$ under RHIC collision energy was plotted in comparison with PHENIX data [19]. As can be seen in Fig.1 the agreement between the result calculated with new model (dash line) and the experiment data is very good, both in shape and magnitude, over a very wide range in y, while a slight disagreement is observed below $y<-2$, where nuclear absorption is expected to play a role. Comparing with the new model, the result calculated with classic model (solid line) has bigger error obviously, which indicated that our new model is more suitable to explain the cold nuclear matter effect in J/Psi production at RHIC. In Fig.1 we also plot the results calculated with energy loss plus dynamical shadowing (dot line) and that of energy loss plus saturation effect (dash do line). The dot line is above the dash dot line at forward rapidity, which indicated that, comparing with the interaction between partons and nuclear background chromo-magnetic field, gluon wave function overlapping play a more important role on the production suppression in the corresponding momentum fraction range.

In Fig.2 the $x_F$ distribution of nuclear modification factor $R_{W/Be}$ under E866 collision energy was plotted. Here the dash line represents result calculated with new model while the solid line represents that of classic model. As can be seen from Fig.2, the agreement between dash line and data [20] is excellent both in shape and magnitude over a wide range, while an obvious disagreement is observed above $x_F>0.7$, where quark-induced processes come into play dominantly. In Fig.2 we also plot the results calculated with energy loss plus dynamical shadowing (dot line) and that of energy loss plus saturation effect (dash dot line). Both of the two lines are similar over the whole $x_F$ range, which hint a same importance of the two effects in this momentum fraction range. Because the $x_2$ detected in E866 is smaller than that in RHIC, no matter the interaction between partons and nuclear background chromo-magnetic field or the gluon wave function overlapping is less strong than that in RHIC. As a result, the difference between dash line and dot line (as well as dash dot line) at E866 is smaller than that at RHIC.

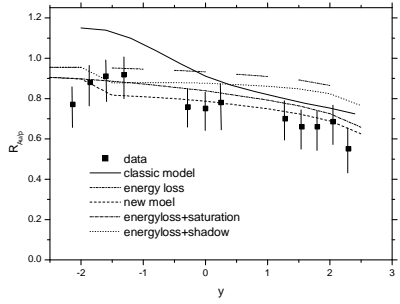 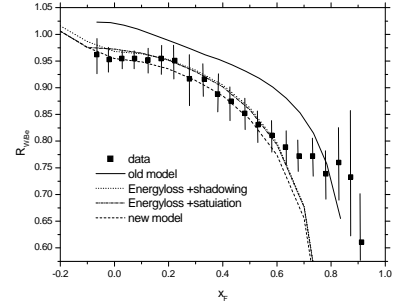

Fig.1. Rapidity distribution of $R_{Au/p}$ at PHENIX.  Fig. 2. $x_F$ distribution of $R_{W/Be}$ at E866.

In Fig.3 the $x_F$ distribution of nuclear modification factor $R_{w/c}$ under HERA collision energy was plotted. As can be seen from the Fig.3, the line calculated with new model reproduce the data [21] excellently both in shape and magnitude, which is much better than that done with the classic model - the later only reproduce the magnitude of the data.

Obviously, comparing with the classic model, the new model is able to reproduce both magnitude and shape of the suppression in p–A collisions under different collision energy more accurately, so it should be more suitable to explain the cold nuclear matter effects in J/Psi production.

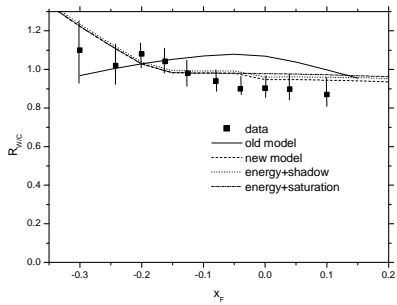 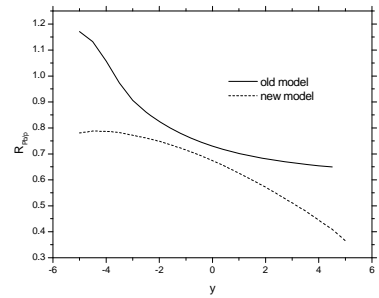

Fig. 3. $x_F$ distribution of $R_{w/c}$ R at HERA.  Fig. 4. Rapidity distribution of $R_{Pb/p}$ at LHC.

Last, the rapidity dependence of J/Psi suppression in p–Pb collisions at the LHC is shown in Fig. 4. Here the dash line represents result calculated with new model while the solid line represents that of classic model. Both of the two lines are similar in shape and magnitude, except that the dash line is lower than the solid line. As a prediction, our result needs to be tested with the future experiment data.

In summary, the cross section ratio of J/Psi produced in heavy ion collisions were calculated with new nuclear effects model in this paper, which model includes such cold nuclear matter effects as the energy loss effect, isospin effect, Cronin effect, gluon saturation effect and dynamical shadowing effect. The advantage of this new model is that all CNM effects have clear physical origin, mostly centered on the idea of multiple parton scattering. It is proved that this new model is able to reproduce accurately almost all existing J/Psi measurements in p–A collisions. In particular the dependence of J/Psi suppression on $x_F$ /y, is well accounted by this model for various atomic mass A and center-of-mass energies. These results eventually verified the availability of the new cold nuclear matter effect model in explaining the cold nuclear matter effects in J/Psi production. As a perspective, we will investigate CNM effects with this new model on other particles production in heavy-ion collisions.

We are very grateful to N. Liu for fruitful discussions. This work is supported by the National Natural Science Foundation of China (No. 10575028).

________________